\RequirePackage{ifpdf}
\ifpdf 
\documentclass[pdftex]{sigma}
\else
\documentclass{sigma}
\fi

\begin{document}

\numberwithin{equation}{section}

\allowdisplaybreaks

\renewcommand{\PaperNumber}{074}

\FirstPageHeading

\renewcommand{\thefootnote}{$\star$}

\ShortArticleName{Vadim Kuznetsov. Informal Biography by Eyes of His First Adviser}

\ArticleName{Vadim Kuznetsov.
Informal Biography \\ by Eyes of His First Adviser\footnote{This paper is a
contribution to the Vadim Kuznetsov Memorial Issue `Integrable
Systems and Related Topics'. The full collection is available at
\href{http://www.emis.de/journals/SIGMA/kuznetsov.html}{http://www.emis.de/journals/SIGMA/kuznetsov.html}}}

\Author{Igor V. KOMAROV}

\AuthorNameForHeading{I.V. Komarov}

\Address{St.~Petersburg State University, St.~Petersburg, Russia}
\Email{\href{mailto:komarov@pobox.spbu.ru}{komarov@pobox.spbu.ru}}

\ArticleDates{Received May 20, 2007; Published online June 22, 2007}

 \Abstract{The paper is dedicated to the memory of prominent theoretical physicist and
mathematician  Dr.~Vadim Kuznetsov who worked, in particular, in the f\/ields of the
nonlinear dynamics, separation of variables, integrability theory, special functions. It includes
 his short research biography, an account of the start of his research career and the list of publications.}


\Classification{01A70}

\renewcommand{\thefootnote}{\arabic{footnote}}
\setcounter{footnote}{0}

Vadim Kuznetsov was a classical ``self-made man'', nobody of his family
ever earned their living by intellectual work. Vadim's talent and will was the
main source of his achievements and authority among the professionals.

The external canvas of his biography looks streamlined: the
Leningrad University, the post-graduate course at the same
university, defence of the thesis and work of a researcher. In
April 1993 Vadim got his f\/irst post-doctoral position (Department
of Mathematics, University of Amsterdam, a NWO post-doctoral
research fellow involved in the project ``Special functions and
quantum inverse scattering method'' supervised by Prof.\
T.H.~Koornwinder). In 1996 V.~Kuznetsov spent six months at the
Technical University in Lyngby, Denmark, where he lectured in
classical and quantum integrable systems. That was followed by two
months of work at the Centre de recherches math\'ematiques,
Universit\'e de Montr\'eal as a research fellow. In 1996 Vadim
Kuznetsov f\/inally settled in Leeds (UK) were he worked at the
university, f\/irst on the grant basis and then at a permanent
position.

In 1980 Vadim Kuznetsov entered the Department of
Physics of the Leningrad University, where since 1982 he had
chosen the major at the department of quantum mechanics. I was
a~lecturer then at the same department, and I had to give a course
with a strange title ``Introduction to Profession''. The students
had not studied quantum mechanics at that moment, and it was not
clear what should be taught. The administration told me to do what
I wished to. I~have chosen a system of talks based on book then
recently published in Russian by W.~Miller ``Separation of
Variables''. Some of the students were given papers in other
topics. Vadim got a paper from {\it Journal of the Optical Society
of America}
 on halo ef\/fect at atmospheric light scattering on ice
crystals. With respect to the talk, Vadim showed persistence in
studying of mathematical and physical details of the paper. Then I
got an impression of him as an accurate person who is good in
laborious tasks. Later I asked Vadim whether this course had any
inf\/luence upon him. The answer was negative. Maybe because of this
course Vadim did not feel any ``trepidation'' with respect to the
classical theory of separation of variables.

 In around a year Vadim overtook me in a stairway and asked for a topic for
 research. On my way I suggested him to f\/ind a semi-classical spectrum of
the  Kowalevski top. I had two ideas~-- to use Kolosov's
projection of the Kowalevski problem
 onto a f\/lat potential problem and to apply the adiabatic
switching  method for f\/inding of the spectrum. I~believed then
 that it was a sure-win problem needing only labor input.
We started regular and intense discussions on this topic. In a few
months Vadim brought me integrals of action calculated in line
with the Kolosov's method. I asked him to investigate limiting
cases for weak and strong f\/ields. In one of the limits one degree
of freedom was disappearing. The formulas were cumbersome, and I
believed that to be a result of some technical mistake.  We got stuck at that
place for several months. Gradually it became clear
that the reason for the problem is that the Kolosov's transformation
where time also transforms is not canonical. Vadim Kuznetsov found
an indication how to proceed correctly in the paper by
S.P.~Novikov and A.P.~Veselov: f\/irst of all, Kowalevski variables
should have been transformed into the Poisson commuting ones. We
invented the next step ourselves. It was restoration of the
canonical variables when the  integrals of motion in the
Lagrangian variables were known. For me and probably for Vadim
that was a very happy moment. A student in his very f\/irst work
started to generate professional ideas at the serious level and
got assured in his research capacities.

Approximately at the same time E.K.~Sklyanin included the
Goryachev--Chaplygin gyrostat I have studied into the framework
of the quantum inverse scattering method (QISM). The Sklyanin's
method had various names -- ``the functional Bethe ansatz'', ``the
magical recipe''. It was obvious that it dealt with f\/inding of the
separation of variables (SoV) in the QISM. I~brought
V.B.~Kuznetsov and his classmate A.V.~Tsiganov to L.D.~Fadeev's
seminar at the Leningrad Branch of Steklov Mathematical Institute
and introduced them to E.K.~Sklyanin. V.~Kuznetsov thoroughly
studied each new paper by E.K. Sklyanin. In particular, some time
 later Vadim studied the algebra of the ref\/lection equations
 (we called it then QISM II), and he was the f\/irst to
construct the separation for one example of QISM II.

At that time Vadim needed to complete technically his education
and to defend a master thesis on the basis of the results already
obtained. It became clear to us that development and application
of the separation of variables in the QISM is a priority task. It
was much later when in the paper  \cite{43} ``Kowalevski top
revisited'' in 2002 Vadim got back to the top problem and found
explicit and quite cumbersome expressions for $2\times 2$ of the
Lax matrix for the Kowalevski top. The results of his master
thesis were published in \cite{1} in {\it Theoretical and Mathematical
Physics} in 1987.

\looseness=-1 In 1986--1989 Vadim was a post-graduate student at
the department of computational physics to which I moved some time
before. Our meetings were regular and fruitful, Vadim was quickly
becoming a qualif\/ied and independent researcher. I set some
problems for him, and Vadim found some problems himself. We also
discussed strategic issues -- in what direction it is preferable
to continue  research in future. Vadim responded positively to
extension of the list of the integrable systems, incorporation of
the known integrable systems into the QISM schemes, accumulation
of the variable separation experience in particular cases. On the
opposite, Vadim rejected my suggestion to look inside the QISM for
ef\/f\/icient algorithms for f\/inding  the spectrum of the integrals of
motion  that were naturally related to the QISM. Among the
``consumers'' of the new ideas of the SoV method, the one mentioned
most often was the theory of special functions that we perceived in
the spirit of the three-volume book by Bateman and Erdely.

Several Lax-like representations for the Kowalevski
problem appeared in the 1980-ies. Vadim rewrote  the available
direct derivation of the separated equations for the case
$so(4,\mathbb C)$ in the spirit of Heine--Horozov. At the same
time A.G.~Reiman and M.A.~Semenov-Tyan-Shanskii constructed
$4\times 4$ and $5\times 5$ Lax matrices for the original
Kowalevski problem. The new Lax matrices depreciated the previous
variants and allowed simplif\/ication of the classical equations of
motion in terms of Prym $\theta$-functions. The attempts to f\/ind
separation of variables from the new Lax matrices did not lead to
any success. As Vadim pointed out later \cite{43} (Kowalevski top
revisited, 2002), until then there was only one separation of
variables known for this problem following from original
Kowalevski's papers. Vadim had identif\/ied in a $4\times 4$ minor
of the Lax matrix for the Kowalevski problem a new $3\times 3$ Lax
matrix for the Goryachev--Chaplygin gyrostat. With A.I.~Bobenko
they had applied this matrix for integration of the respective
equations of motion \cite{3}.

After the post-graduate course and defence of his thesis (referees
E.K.~Sklyanin and A.R.~Its) earlier than the scheduled date in
October of 1989 V.~Kuznetsov worked as a researcher at the
Department of Computational Physics.

\looseness=1
In autumn of 1991 Ernie Kalnins that worked for many years on the
coordinate separation of variables visited our university. Ernie
wanted to see Riga, and Vadim accompanied him on that trip. Before
Ernie Kalnins's visit Vadim wrote on his own initiative several
papers on co-ordinate separation of variables for free motion in
the spaces of permanent curvature in classical and quantum
mechanics on the basis of  currents algebra \cite{9, 10, 11}. The most
important was that it was possible not only to reproduce quite
transparently the known results on co-ordinate separation of
variables for the Laplace operators in the low-dimensional spaces
of constant curvature from the Gaudin algebra, but also to extend
this procedure to the spaces with arbitrary number of dimensions
and to construct reduction of quadrics. These results attracted
the considerable interest at the conference in Obninsk of N.Ya.~Vilenkin
who asked questions to Vadim for quite a long time and in
great detail. Later (February 25, 1993) Vadim received for this
cycle of papers the Award of the Academiae Europaeae (London),
instituted by the club of the Russian members of the Academiae for
young Russian scientists.

The end of the 1980-ies and the beginning of the 1990-ies were a
dif\/f\/icult time for survival. The USSR economy was collapsing fast.
The researcher's salary was evidently insuf\/f\/icient to support the
family. Vadim found odd jobs, up to petty trade near metro
stations that was very humiliating for him. Vadim found a
postdoctoral position at the University of Amsterdam. I~learned
about his decision to go abroad when Vadim asked me for a letter
of recommendation. In Amsterdam Vadim worked together with the
expert on the special functions theory Thom H. Koornwinder. The
results of their joint research were included into the 1994 papers
\cite{18, 25} were properties of the Gauss hypergeometric function (in
particular, contiguous function relations) were derived in a
regular way from the properties of the $R$-matrix algebra. The
papers of this period developed the ideas of the separation of
variables for rational and trigonometric linear and quadratic
$R$-matrix algebra.

The list of Vadim's co-authors was extending fast. Vadim advanced
his knowledge and mastered new f\/ields of mathematics and theoretical physics.

In 1994 a many-year co-operation of V.B.~Kuznetsov and
E.K.~Sklyanin had started. They obtained separation of variables
for the $A\sb 2$ Jack polynomials. Systematic research in the
f\/ield of the Ruijsenaars--Schneider model was also started
\cite{26}~(the co-authors were F.W.~Nijhof\/f, O.~Ragnisco,
E.K.~Sklyanin). I would like to mention here the observation on
the relation of the Bethe equations with the integrable
time-discretisation of the model.

It is known that for the separation of variables by means of the
poles of the Baker--Akhiezer function $\bf f$ its special
normalisation is needed. For many examples the scalar product
$({\bf f, a})$ with a constant vector
 ${\bf a}$ was suf\/f\/icient, but sometimes the number of poles
 exceeded the number of degrees of freedom. In the paper on
 ${\cal D}_n$ type periodic Toda
lattice V.~Kuznetsov managed to f\/ind a correct normalising vector
$\bf a$ depending on the dynamic variables \cite{30}. But that was his
last work involving poles of the Baker--Akhiezer function.
Presence of artif\/icial tricks made Vadim to look for more
universal approaches.

The papers by V.B.~Kuznetsov and E.K.~Sklyanin  \cite{32} developed a
new more  systematic approach to separation of variables and
presented relation of the SoV with the B\"añklund transformation.
Invention of the ``spectrality'' property was a key point. The
Baxter $Q$-operator became a subject of attention of many
researchers. Vadim often said that a new understanding of the
separation of variables started for him from the 1998 paper \cite{32}
``On B\" acklund transformations for many-body systems''.
Development of this technique by the group of mathematicians with
participation of V.~Kuznetsov led to qualitatively new results on
factorisation of the symmetric Jack, Hall--Littlewood and
Macdonald polynomials.

For more than 10 years Vadim maintained research contacts with the
group of Mark Adler, Pierre van Moerbeke and Pol Vanhaecke, with
whom he published only two papers \cite{40, 41} on geometric aspects of
the B\"acklund transform.

In 2000 Vadim organised jointly with Frank Nijfof\/f the
International Workshop on Mathematical Methods of Regular Dynamics
dedicated to the 150th anniversary of Sonya Kowalevski. In 2002
Vadim Kuznetsov and Frank Nijhof\/f edited the materials of the
Workshop \cite{42}. In 2003 Vadim was an organiser of the workshop in
Edinburgh that was dedicated to the classical works by H.~Jack,
P.~Hall, D.E.~Littlewood and I.G.~MacDonald on Symmetric Functions
and their relation to Representation Theory of Symmetric Groups.
The materials of this workshop edited by V.~Kuznetsov and S.~Sahi
\cite{51} were published after his death. In that issue Brian
D.~Sleeman and Evgeny K.~Sklyanin published a paper dedicated to
Vadim Kuznetsov's memory\footnote{Sleeman B.D., Sklyanin E.K.,
Vadim Borisovich Kuznetsov 1963--2005, {\it Contemp. Math.} {\bf
417} (2006), 357--360.}.

Vadim worked a lot directly with formulas. Sometimes he obtained
results like ``Hidden symmetry of the quantum Calogero--Moser
system'', 1996, that at  the f\/irst glance had no relation to the
general theory. This list can be extended and it may give start to
future studies.

I visited Vadim and his family in Amsterdam, Copenhagen and Leeds,
and we also met during Vadim's short visits to St.~Petersburgh and
Dubna. In July 1998 both of us were in dif\/ferent regions of
Germany and specially came to Kaizerslautern to see each other. We
were walking in the city where some Sunday German Fest was
bustling, and Vadim enthusiastically spoke about new ideas in the
joint publications with E.K.~Sklyanin. Later the research aspect
of our discussions moved to a background. During our last meetings
in Leeds in May 2005 Vadim looked very tired, he has immense
teaching  and administrative load. We tried to talk about
mathematics without  big  success. Vadim mentioned that he
intended to start a completely new research area  not related to
separation of variables. I know nothing about implementation of
these plans.

Vadim was survived by his wife Olga and son Simon.


\pdfbookmark[1]{Dr. Vadim B. Kuznetsov:  List of
publications}{ref}

\renewcommand{\refname}{Dr. Vadim B. Kuznetsov.  List of publications}
\LastPageEnding


\end{document}